\documentclass[preprint,12pt]{elsarticle}

\usepackage{amssymb}

\makeatletter
\newcommand{\vast}{\bBigg@{4}}
\makeatother
\usepackage{multirow}
\usepackage{graphicx}
\usepackage{epstopdf}
\usepackage[cmex10]{amsmath}
\usepackage{amssymb}

\usepackage{multirow}
 \usepackage{algpseudocode}
\usepackage{algorithm}
\usepackage{amsthm}
\theoremstyle{definition}
\theoremstyle{remark}

\newtheorem{rem}{Remark}

\makeatletter
\g@addto@macro\th@remark{\thm@headpunct{\normalfont:}}
\makeatother
\makeatletter
\newcommand{\distas}[1]{\mathbin{\overset{#1}{\kern\z@\sim}}}%
\newsavebox{\mybox}\newsavebox{\mysim}
\newcommand{\distras}[1]{%
  \savebox{\mybox}{\hbox{\kern3pt$\scriptstyle#1$\kern3pt}}%
  \savebox{\mysim}{\hbox{$\sim$}}%
  \mathbin{\overset{#1}{\kern\z@\resizebox{\wd\mybox}{\ht\mysim}{$\sim$}}}%
}

\journal{International Journal of Electronics and Communications}

\begin{document}

\begin{frontmatter}



\title{Cooperative Relaying in Underlay Cognitive Systems with Hardware Impairments}


\author[label1,label2]{Nikolaos~I.~Miridakis}
\ead{nikozm@unipi.gr}
\author[label3]{Dimitrios~D.~Vergados}
\ead{vergados@unipi.gr}
\author[label2]{Angelos~Michalas}
\ead{amichalas@kastoria.teikoz.gr}
\address[label1]{Department of Computer Engineering, Piraeus University of Applied Sciences, 122 44, Aegaleo, Greece}

\address[label2]{Department of Informatics and Computer Technology, Technological Education Institute of Western Macedonia, 52 100, Kastoria, Greece.}
\address[label3]{Department of Informatics, University of Piraeus, 185 34, Piraeus, Greece.}

\begin{abstract}
The performance of an underlay cognitive (secondary) dual-hop relaying system with multiple antennas and hardware impairments at each transceiver is investigated. In particular, the outage probability of the end-to-end (e2e) communication is derived in closed-form, when either transmit antenna selection with maximum ratio combining (TAS/MRC), or TAS with selection combining (TAS/SC) are established in each hop. Simplified asymptotic outage expressions are also obtained, which manifest the diversity and array order of the system, the effectiveness of the balance on the number of transmit/receive antennas, and the impact of hardware impairments to the e2e communication.
\end{abstract}


\end{frontmatter}


\section{Introduction}
Underlay cognitive transmission represents one of the most promising spectrum sharing techniques, where secondary (unlicensed) users utilize the spectrum resources of another primary (licensed) service \cite{ref9,ref99}. To this end, the transmission power of the cognitive system is limited, such that its interference onto the primary users remains below prescribed tolerable levels. However, this dictated constraint dramatically affects the coverage and/or capacity of the secondary communication. Such a condition can be effectively counteracted with the aid of relayed transmission \cite{ref10}. On the other hand, using multiple antennas at each node (and, thus, benefiting from the emerged spatial diversity of each transmitted stream) is another effective approach to enhance the performance of a cognitive system \cite{ref4}-\cite{refextra}. Due to the complementary benefits of relayed transmission and spatial diversity gain, the performance analysis of these schemes, under the cognitive transmission regime, is of prime interest lately (e.g., see \cite{ref9,ref11,ref12} and references therein).

All the previous research works assumed ideal hardware at the transmitter and/or receiver end, where the scenario of impaired transceivers (namely, non-ideal hardware) was neglected. Nevertheless, this condition represents a rather overoptimistic scenario for practical applications. More specifically, the hardware gear of wireless transceivers may be subject to impairments, such as I/Q imbalance, phase noise, and high power amplifier nonlinearities \cite{ref222}. Yet, very few research works have analytically investigated the impact of hardware impairments. Specifically, the outage probability of one-way \cite{ref1} and two-way \cite{ref7} dual-hop relayed transmission systems is analytically expressed, in the case of single-antenna transceivers and non-cognitive environments. Nevertheless, a corresponding analysis when multiple antennas are employed at each node and/or under a cognitive transmission regime lacks from the open technical literature so far. Capitalizing on these observations, in this paper, the performance of a dual-hop cognitive system with multiple-antennas and hardware impairments at the transceiver of each hop is analytically investigated. Particularly, two popular spatial diversity techniques are adopted, namely, transmit antenna selection with maximum ratio combining (TAS/MRC), or TAS with selection combining (TAS/SC) are established in each hop. It is noteworthy that with TAS, the number of RF chains required is equal to the number of antennas selected for communication, a rather cost-efficient solution for various applications. Most importantly, even with lower complexity, TAS gives full diversity. Hence, it is recently of great importance due to its low feedback demand \cite{ref4}-\cite{ref12}, whereas it plays a key role to the uplink of 4G networks \cite{extraaa}. Also, the decode-and-forward (DF) protocol is used for the relayed transmission.

New closed-form expressions are derived with respect to the outage probability of the system, which generalize some previously reported results. In addition, simplified asymptotic expressions of the outage probability, in the high signal transmission regime are also obtained, revealing the diversity and array order of the system, the effectiveness of the balance on the number of transmit/receive antennas, and the impact of hardware impairments to the end-to-end (e2e) communication. Specifically, it is demonstrated that the diversity order and the performance difference between TAS/MRC and TAS/SC remain unaffected from hardware impairments.

\section{System Model}
\begin{figure}[!t]
\centering
\includegraphics[keepaspectratio,width=5.0in]{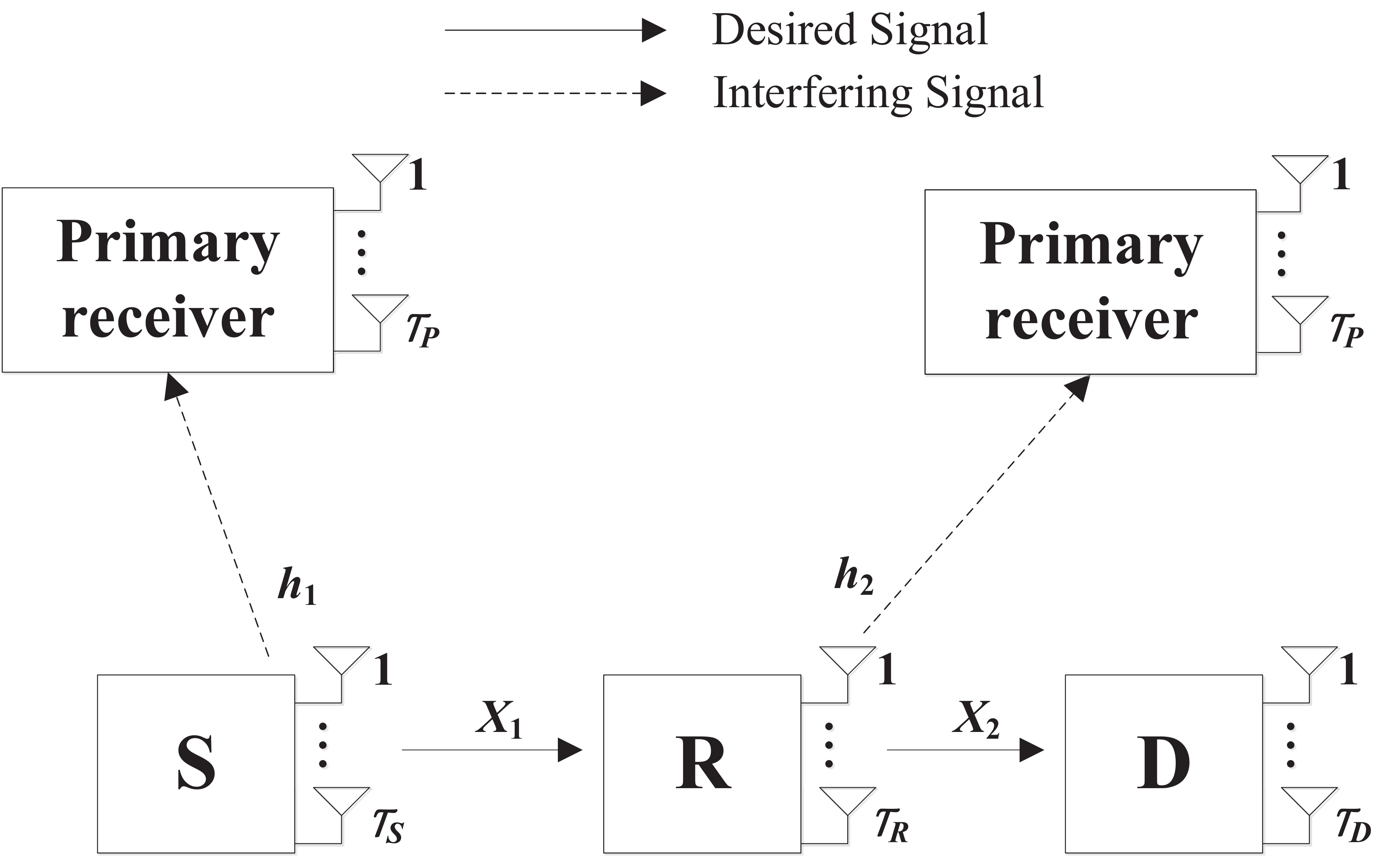}
\caption{The considered system configuration and the channel fading gains which are reflected at the primary receivers, relay and secondary receiver.}
\label{fig1}
\end{figure}

Consider an underlay secondary system, where a source node (S) communicates with a destination node (D) via an intermediate relay node (R), as illustrated in Fig. \ref{fig1}. The direct link between S and D is assumed to be absent due to strong propagation attenuation and keeping in mind that the transmission power of the underlay system is, in principle, maintained quite low. Let $\mathcal{T}_{\text{P}}$, $\mathcal{T}_{\text{S}}$, $\mathcal{T}_{\text{R}}$ and $\mathcal{T}_{\text{D}}$ denote the number of antennas at the primary receivers, S, R and D, respectively. The antennas of each node are sufficiently separated from one another (with respect to the transmission wavelength) to prevent any channel fading correlation.\footnote{The presented results can serve as upper performance bounds for the more general case of correlated fading.}
Also, assume identical Rayleigh faded channels for the antennas of each node and not necessarily identical channels from one node to another. Furthermore, the DF transmission protocol is adopted for the dual-hop relayed transmission, which has shown very good results in terms of error rates and outage performance \cite{ref8}. For mathematical tractability, let $\left\{M_{1},N_{1}\right\}\triangleq\left\{\mathcal{T}_{S},\mathcal{T}_{R}\right\}$ for the 1st hop, while $\left\{M_{2},N_{2}\right\}\triangleq\left\{\mathcal{T}_{R},\mathcal{T}_{D}\right\}$ for the 2nd hop. 

In the case when signal distortion due to hardware impairments is present and TAS is established at the transmitter side, the input-output relation of the received signal at the $i$th transmission hop is given by $\textbf{y}_{i}=p_{i}\textbf{g}_{i}(s_{i}+e_{T_{i}})+\textbf{e}_{R_{i}}+\textbf{v}_{i}$. Let $\{\textbf{y}_{i},\textbf{g}_{i},\textbf{e}_{R_{i}},\textbf{v}_{i}\} \in \mathbb{C}^{N_{i}\times 1}$ denote vectors for the received signal, the channel fading, the distortion noise to the received signal, and the additive white Gaussian noise (AWGN) of the $i$th hop, respectively. Moreover, $p_{i}$, $s_{i}$ and $e_{T_{i}}$ (scalar values) correspond to the transmission power, the transmitted signal and the distortion noise to the transmitted signal of the $i$th hop, respectively. Notice that $\textbf{v}_{i}\overset{}{\sim} \mathcal{CN}(0,N_{0}\textbf{I}_{N_{i}})$, $e_{T_{i}}\overset{}{\sim}\mathcal{CN}(0,p_{i}\kappa^{2}_{T_{i}})$ and $\textbf{e}_{R_{i}}\overset{}{\sim} \mathcal{CN}(0,\kappa^{2}_{R_{i}}N_{i}\textbf{I}_{N_{i}})$ \cite{ref1}, \cite{ref11111}. Also, $\kappa_{T_{i}}$ and $\kappa_{R_{i}}$ represent certain parameters describing the underlying distortion noise at the transmitter and receiver, respectively, $N_{0}$ is the power spectral density of AWGN, while $\textbf{I}_{n}$ stands for the $n$-sized identity matrix. Further, assume that the hardware quality of each antenna is identical for the same device, i.e., there is an equal distortion noise variance per antenna for each node, while this variance is perfectly known at the receiver side (e.g., through pilot or feedback signaling). In principle, channel state information (CSI) of the links between the primary and secondary nodes can be obtained through a feedback channel from the primary service or through a band manager that mediates the exchange of information between the primary and secondary networks \cite{ref9,ref10}.

Hence, the received signal-to-noise-and-distortion ratio (SNDR) at the $i$th hop reads as \cite[Eq. (6)]{ref1}
\begin{equation}
\text{SNDR}_{i}=\frac{p_{i}\mathcal{X}_{i}}{\left(p_{i}\kappa_{T_{i}}^{2}\mathcal{X}_{i}\right)+n_{i}}, \ \ i \in \left\{1,2\right\},
\label{sndr}
\end{equation} 
where $n_{i}\triangleq \kappa^{2}_{R}N_{i}+N_{0}$ and $\mathcal{X}_{i}$ is the channel fading gain of the desired signal, which differs according to the adopted reception strategy (which is analyzed in the next section). Notice that $p_{i}$ can not exceed the predefined interfering power threshold to the primary receiver (the so-called interference temperature), such that $p_{i}\triangleq \min \left\{p_{\text{max}},w_{i}/h^{2}_{i}\right\}$, where $h^{2}_{i}$, $w_{i}$ and $p_{\text{max}}$ denote the aggregated channel fading gain between the secondary transmitter (S or R, depending on the hop) and the primary receiver (i.e., $h^{2}_{i}\triangleq \sum^{\mathcal{T}_{\text{P}}}_{l=1}h^{2\:(l)}_{i}$), the tolerable interfering power threshold and the maximum achievable transmission power, respectively. Hence, (\ref{sndr}) becomes
\begin{equation}
\text{SNDR}_{i}= \left\{
\begin{array}{c l}     
    \frac{p_{\text{max}}\mathcal{X}_{i}}{\left(p_{\text{max}}\kappa_{T_{i}}^{2}\mathcal{X}_{i}\right)+n_{i}}, & h^{2}_{i} < \frac{w_{i}}{p_{\text{max}}},\\
    \frac{w_{i}\mathcal{X}_{i}/h^{2}_{i}}{\left(w_{i}\kappa_{T_{i}}^{2}\mathcal{X}_{i}/h^{2}_{i}\right)+n_{i}}, & h^{2}_{i} > \frac{w_{i}}{p_{\text{max}}}.
\end{array}\right.
\label{sndrr}
\end{equation}

\section{Performance Analysis}
We start by defining the cumulative distribution function (CDF) of SNDR at the $i$th hop, namely $F_{\text{SNDR}_{i}}(.)$. Based on (\ref{sndrr}), we have that $F_{\text{SNDR}_{i}}(\gamma)=\int^{\infty}_{0}F_{\text{SNDR}_{i}|h^{2}_{i}}(\gamma|h)f_{h^{2}_{i}}(h)dh$, yielding
\begin{align}
\nonumber
&F_{\text{SNDR}_{i}}(\gamma)=F_{\mathcal{X}_{i}}\left(\frac{n_{i}\gamma}{p_{\text{max}}(1-\kappa^{2}_{T_{i}}\gamma)}\right)\left(\frac{\Gamma_{L}(\mathcal{T}_{P},\frac{w_{i}}{p_{\text{max}}\overline{y}_{i}})}{\Gamma(\mathcal{T}_{P})}\right)\\
&\ \ \ \ +\int^{\infty}_{\frac{w_{i}}{p_{\text{max}}}}F_{\mathcal{X}_{i}}\left(\frac{n_{i}\gamma h}{w_{i}(1-\kappa^{2}_{T_{i}}\gamma)} \right)\frac{h^{\mathcal{T}_{\text{P}}-1}\exp{\left(-\frac{h}{\overline{y}_{i}}\right)}}{\Gamma(\mathcal{T}_{\text{P}})\overline{y}_{i}^{\mathcal{T}_{\text{P}}}}dh,
\label{4b}
\end{align} 
where $\Gamma(.)$ and $\Gamma_{L}(.,.)$ are the Euler's Gamma function \cite[Eq. (8.310.1)]{tables} and the lower incomplete Gamma function \cite[Eq. (8.350.1)]{tables}, respectively. Moreover, $f_{h^{2}_{i}}(h)\triangleq h^{\mathcal{T}_{\text{P}}-1}\exp{\left(-h/\overline{y}_{i}\right)}/(\Gamma(\mathcal{T}_{\text{P}})\overline{y}_{i}^{\mathcal{T}_{\text{P}}})$ is the probability density function (PDF) of $h^{2}_{i}$ with $\overline{y}_{i}$ representing the average channel fading gain of $h^{2}_{i}$, $F_{\mathcal{X}_{i}}(.)$ is the CDF of $\mathcal{X}_{i}$, and $\text{Pr}[.]$ stands for the probability operator. Also, (\ref{4b}) is obtained utilizing \cite[Eq. (3.351.1)]{tables} and using the fact that $\mathcal{X}_{i}$ and $h^{2}_{i}$ are mutually independent random variables. Notice that the condition $\kappa^{2}_{T_{i}} < 1/\gamma$ should be satisfied, which is usually the case in most practical scenarios, since typical values of $\kappa_{T}$ lie in the range of [$0.08, 0.175$] \cite{ref6}.

Meanwhile, the desired signal power is expressed as\footnote{The clear-optimal scheme would operate with respect to SNDR \cite{ref4}. Herein, we consider the suboptimal (yet more efficient in terms of signaling overhead) scheme operating with respect to the channel gains of the desired signal.}
\begin{align*}
\mathcal{X}_{i}=\left\{\max_{\substack{j}}\sum^{N_{i}}_{l=1}x^{(j,l)}_{i}\right\}^{M_{i}}_{j=1}
\end{align*}
for TAS/MRC or
\begin{align*}
\mathcal{X}_{i}=\left\{\max_{\substack{j,l}}x^{(j,l)}_{i}\right\}^{M_{i},N_{i}}_{j=1,l=1}
\end{align*}
for TAS/SC, where $x^{(j,l)}_{i}$ represents the channel power fading gain of the desired signal from the $j$th transmit to the $l$th receive antenna at the $i$th hop.  

\subsection{Preliminary Statistics}
In the case of TAS/MRC, the CDF of $\mathcal{X}_{i}$ yields as
\begin{align}
\nonumber
F_{\mathcal{X}_{i}}(z)&=\left(1-\frac{\Gamma_{U}(N_{i},z/\overline{x}_{i})}{\Gamma(N_{i})}\right)^{M_{i}}\\
&=\Xi(M_{i},N_{i})(z/\overline{x}_{i})^{\phi}\text{exp}\left(-\frac{k z}{\overline{x}_{i}}\right),\ \ \ \ z\geq 0,
\label{cdfxi}
\end{align}
where $\Gamma_{U}(.,.)$ denotes the upper incomplete Gamma function \cite[Eq. (8.350.2)]{tables}, while
\begin{align}
\nonumber
\Xi(M_{i},N_{i})\triangleq &\sum^{M_{i}}_{k=0}\sum^{k}_{p_{1}=0}\sum^{p_{1}}_{p_{2}=0}\cdots\sum^{p_{N_{i}-2}}_{p_{N_{i}-1}=0}\frac{\binom{M_{i}}{k}(-1)^{k}k!}{p_{N_{i}-1}!}\\
& \times \prod^{N_{i}-1}_{t=1}\left[\frac{1}{(p_{t-1}-p_{t})!(t!)^{p_{t}-p_{t+1}}}\right],
\label{Xi}
\end{align}
and $\phi=\sum^{N_{i}-1}_{q=0}p_{q}$, $p_{N_{i}}=0$, and $p_{0}=k$. For the closed-form derivation of (\ref{cdfxi}), first \cite[Eqs. (8.352.4), (1.111)]{tables} and then \cite[Eq. (A.3)]{ref2} are utilized. 

In the case of TAS/SC, the CDF of $\mathcal{X}_{i}$ is given in this case by
\begin{align*}
F_{\mathcal{X}_{i}}(z)=\sum^{N_{i} M_{i}}_{k=0}\binom{N_{i} M_{i}}{k}(-1)^{k}\text{exp}\left(-\frac{k z}{\overline{x}_{i}}\right).
\end{align*}
Notably, the latter expression can be directly obtained from (\ref{cdfxi}), by setting $\Xi(M_{i},N_{i})\triangleq \sum^{N_{i}M_{i}}_{k=0}\binom{N_{i}M_{i}}{k}(-1)^{k}$ and $\phi=0$. Thus, in the rest of this paper, (\ref{cdfxi}) will be used to denote both the TAS/MRC and TAS/SC scenarios, correspondingly.

\subsection{Outage Probability of the e2e Communication}
Using (\ref{cdfxi}) in (\ref{4b}), invoking \cite[Eq. (3.381.3)]{tables} and after some algebra, the closed-form CDF of SNDR for the $i$th hop is expressed as
\begin{align}
\nonumber
F_{\text{SNDR}_{i}}(\gamma)&=\Xi(M_{i},N_{i})\left(\frac{n_{i}\gamma}{(1-\kappa^{2}_{T_{i}}\gamma)\overline{x}_{i}}\right)^{\phi}\\
&\times \vast\{\frac{\left(\frac{\Gamma_{L}(\mathcal{T}_{P},\frac{w_{i}}{p_{\text{max}}\overline{y}_{i}})}{\Gamma(\mathcal{T}_{P})p_{\text{max}}^{\phi}}\right)}{\text{exp}\left(\frac{k n_{i}\gamma}{p_{\text{max}}(1-\kappa^{2}_{T_{i}}\gamma)\overline{x}_{i}}\right)}+\frac{\left(\frac{\Gamma_{U}\left(\phi+\mathcal{T}_{\text{P}},\frac{k n_{i}\gamma}{p_{\text{max}}(1-\kappa^{2}_{T_{i}}\gamma)\overline{x}_{i}}+\frac{w_{i}}{p_{\text{max}}\overline{y}_{i}}\right)}{\Gamma(\mathcal{T}_{\text{P}})\overline{y}_{i}^{\mathcal{T}_{\text{P}}}w_{i}^{\phi}}\right)}{\left(\frac{k n_{i}\gamma}{w_{i}(1-\kappa^{2}_{T_{i}}\gamma)\overline{x}_{i}}+\frac{1}{\overline{y}_{i}}\right)^{\phi+\mathcal{T}_{\text{P}}}}\vast\}.
\label{sndrfinal}
\end{align}
Moreover, the only special functions within (\ref{sndrfinal}), i.e., $\Gamma_{L}(.,.)$ and $\Gamma_{U}(.,.)$, can be easily converted into finite sum series of elementary functions, according to \cite[Eq. (8.352.4)]{tables}, although omitted herein for ease of presentation. It is noteworthy that (\ref{sndrfinal}) coincides with \cite[Eq. (7)]{ref4}, when $\mathcal{T}_{\text{P}}=1$ and $\kappa_{T_{i}}=\kappa_{R_{i}}=0$ (i.e., ideal hardware). 

The outage probability is defined as the probability that the $e2e$ SNDR, i.e., SNDR$_{e2e}$, falls below a certain threshold value, namely $\gamma_{\text{th}}$, such that $P_{\text{out}}(\gamma_{\text{th}})=\text{Pr}\left[\text{SNDR}_{e2e}\leq \gamma_{\text{th}}\right]$. It is straightforward to show that $P_{\text{out}}(\gamma_{\text{th}})\triangleq F_{e2e}(\gamma_{\text{th}})$, where $F_{e2e}(.)$ is the CDF of SNDR$_{e2e}$, which is obtained as
\begin{align*}
F_{e2e}(\gamma_{\text{th}})=F_{\text{SNDR}_{1}}(\gamma_{\text{th}})+F_{\text{SNDR}_{2}}(\gamma_{\text{th}})-F_{\text{SNDR}_{1}}(\gamma_{\text{th}})F_{\text{SNDR}_{2}}(\gamma_{\text{th}}),
\end{align*}
since $\text{SNDR}_{e2e}=\min\left[\text{SNDR}_{1},\text{SNDR}_{2}\right]$ \cite{ref8}. 

\subsection{Special Case: Outage Probability of the e2e Communication for Single-Antenna Systems}
In the case when all the nodes are equipped with single-antennas, $F_{\mathcal{X}_{i}}(.)$ and $f_{h^{2}_{i}}(.)$ reduce to the classical exponential CDF/PDF. Thus, keeping in mind that $F_{e2e}(.)$ can be alternatively derived as $F_{e2e}(\gamma_{\text{th}})=1-(1-F_{\text{SNDR}_{1}}(\gamma_{\text{th}}))(1-F_{\text{SNDR}_{2}}(\gamma_{\text{th}}))$, outage probability simplifies to 
\begin{align}
\nonumber
P_{\text{out}}(\gamma_{\text{th}})=&1-\text{exp}\left(-\sum^{2}_{i=1}\frac{\gamma_{\text{th}} n_{i}}{p_{\text{max}}(1-\gamma_{\text{th}} \kappa^{2}_{i})\overline{x}_{i}}\right)\\
&\times \prod^{2}_{i=1}\left[\frac{\text{exp}\left(-\frac{w_{i}}{p_{\text{max}}\overline{y}_{i}}\right)\gamma_{\text{th}} \overline{y}_{i}n_{i}}{\gamma_{\text{th}} \overline{y}_{i}n_{i}+w_{i}\overline{x}_{i}(1-\gamma_{\text{th}} \kappa^{2}_{i})}\right].
\label{sndrfinal1}
\end{align}
Notice that when there is no transmission power constraint (i.e., $p_{i}=p_{\text{max}}$), (\ref{sndrfinal1}) reduces to \cite[Eq. (31)]{ref1}.

\subsection{Asymptotic Analysis}
The previously derived expressions are exact; albeit they admit a more amenable formulation in the high SNDR regime. 

\subsubsection{$p_{\text{max}}\rightarrow \infty$}
In the asymptotically high SNDR regime, while utilizing \cite[Eq. (3.381.4)]{tables}, (\ref{4b}) becomes (for both TAS/MRC and TAS/SC)
\begin{align*}
F_{\text{SNDR}_{i}|p_{\text{max}}\rightarrow \infty}(\gamma)\approx\int^{\infty}_{0}F_{\mathcal{X}_{i}|p_{\text{max}}\rightarrow \infty}\left(\textstyle \frac{n_{i}\gamma h}{w_{i}(1-\kappa^{2}_{T_{i}}\gamma)}\right)f_{h^{2}_{i}}(h)dh,
\end{align*}
yielding
\begin{align}
F_{\text{SNDR}_{i}|p_{\text{max}}\rightarrow \infty}(\gamma)\approx \frac{\Xi(M_{i},N_{i})\left(\frac{n_{i}\gamma}{w_{i}(1-\kappa^{2}_{T_{i}}\gamma)\overline{x}_{i}}\right)^{\phi}\Gamma(\phi+\mathcal{T}_{\text{P}})}{\left(\frac{k n_{i}\gamma}{w_{i}(1-\kappa^{2}_{T_{i}}\gamma)\overline{x}_{i}}+\frac{1}{\overline{y}_{i}}\right)^{\phi+\mathcal{T}_{\text{P}}}\overline{y}_{i}^{\mathcal{T}_{\text{P}}}},
\label{cdfxasympt}
\end{align}

\subsubsection{TAS/MRC when $p_{\text{max}},\overline{x}_{i}\rightarrow \infty$}
In this case, both the maximum available transmission power and the fading gain of the desired channel are asymptotically high (e.g., very low propagation attenuation). Recognizing that $\Gamma_{U}(\alpha,x)\approx \Gamma(\alpha)-x^{\alpha}/\alpha$ as $x\rightarrow 0^{+}$ \cite[Eq. (8.354.2)]{tables}, (\ref{4b}) is efficiently approached by
\begin{align}
F_{\text{SNDR}_{i}|p_{\text{max}},\overline{x}_{i}\rightarrow \infty}(\gamma)\approx\frac{\left(\frac{n_{i}\gamma }{w_{i}(1-\kappa^{2}_{T_{i}}\gamma)\overline{x}_{i}}\right)^{N_{i} M_{i}}\Gamma(N_{i} M_{i}+\mathcal{T}_{\text{P}})}{N_{i}^{M_{i}}\Gamma(N_{i})^{M_{i}}\overline{y}_{i}^{-(N_{i} M_{i}+\mathcal{T}_{\text{P}})}}.
\label{cdfxasympt1}
\end{align}

\subsubsection{TAS/SC when $p_{\text{max}},\overline{x}_{i}\rightarrow \infty$}
Following similar lines of reasoning as in the TAS/MRC scenario, we have 
\begin{align}
F_{\text{SNDR}_{i}|p_{\text{max}},\overline{x}_{i}\rightarrow \infty}(\gamma)\approx \frac{\left(\frac{n_{i}\gamma }{w_{i}(1-\kappa^{2}_{T_{i}}\gamma)\overline{x}_{i}}\right)^{N_{i} M_{i}}\Gamma(N_{i} M_{i}+\mathcal{T}_{\text{P}})}{\overline{y}_{i}^{-(N_{i} M_{i}+\mathcal{T}_{\text{P}})}}.
\label{cdfxasympt2}
\end{align}

In the above cases, outage probability of the $e2e$ SNDR is approximated by $P_{\text{out}|p_{\text{max}},\overline{x}_{i}\rightarrow \infty}(\gamma_{\text{th}})\approx F_{\text{SNDR}_{1}}(\gamma_{\text{th}})+F_{\text{SNDR}_{2}}(\gamma_{\text{th}})$.

\begin{rem}
Since $P_{\text{out}}\propto \overline{x}_{i}^{-(N_{i} M_{i})}$ for both the TAS/MRC and TAS/SC scenarios, it is straightforward to show that the diversity order of the system remains $\min\left\{\mathcal{T}_{\text{S}} \mathcal{T}_{\text{R}},\mathcal{T}_{\text{R}} \mathcal{T}_{\text{D}}\right\}$, it is maximized when equal number of antennas are used for transmission and reception, and it is not affected by the impact of hardware impairments.\footnote{This is the classical diversity order of cognitive systems without hardware impairments \cite{ref4}.}  
\end{rem}

The performance difference between the two diversity techniques is highlighted in the underlying array order (or coding gain). For identical channel fading statistics of each hop, we have that $\kappa_{T_{1}}=\kappa_{T_{2}}\triangleq \kappa$, $w_{1}=w_{2}\triangleq w$, $n_{1}=n_{2}\triangleq n$ (thus, $\kappa_{R_{1}}=\kappa_{R_{2}}$ and $N_{0_{1}}=N_{0_{2}}$), $\overline{y}_{1}=\overline{y}_{2}\triangleq \overline{y}$ and $\overline{x}_{1}=\overline{x}_{2}\triangleq \overline{x}$. Thus, the array order of the dual-hop system can be expressed as
\begin{align*}
\nonumber
\text{Array Order}_{(\text{TAS/SC})}=&\frac{2 w(1-\kappa^{2}\gamma_{\text{th}})}{n \gamma_{\text{th}} \overline{y}^{1+\frac{\mathcal{T}_{\text{P}}}{\min\left\{\mathcal{T}_{\text{S}} \mathcal{T}_{\text{R}},\mathcal{T}_{\text{R}} \mathcal{T}_{\text{D}}\right\}}}}\\
&\times \Gamma(\min\left\{\mathcal{T}_{\text{S}} \mathcal{T}_{\text{R}},\mathcal{T}_{\text{R}} \mathcal{T}_{\text{D}}\right\}+\mathcal{T}_{\text{P}})^{-\frac{1}{\min\left\{\mathcal{T}_{\text{S}} \mathcal{T}_{\text{R}},\mathcal{T}_{\text{R}} \mathcal{T}_{\text{D}}\right\}}},
\end{align*}
and
\begin{align*}
\nonumber
\text{Array Order}_{(\text{TAS/MRC})}=&\text{Array Order}_{(\text{TAS/SC})}\\ &\times(\Gamma\left(\min\left\{\mathcal{T}_{\text{R}},\mathcal{T}_{\text{D}}\right\}) \min\left\{\mathcal{T}_{\text{R}},\mathcal{T}_{\text{D}}\right\}\right)^{\frac{1}{\min\left\{\mathcal{T}_{\text{R}},\mathcal{T}_{\text{D}}\right\}}}.
\end{align*}

\begin{rem}
The impact of hardware impairments is manifested in the array order of the system, for both diversity scenarios, as expected. Interestingly, the performance gain of the array order for TAS/MRC as compared with TAS/SC is $\propto (\Gamma\left(\min\left\{\mathcal{T}_{\text{R}},\mathcal{T}_{\text{D}}\right\}) \min\left\{\mathcal{T}_{\text{R}},\mathcal{T}_{\text{D}}\right\}\right)^{\frac{1}{\min\left\{\mathcal{T}_{\text{R}},\mathcal{T}_{\text{D}}\right\}}}$, thereby it remains unaffected from hardware impairments (i.e., independent from $\kappa$). 
\end{rem}

\section{Numerical Results and Discussion}
\begin{figure}[!t]
\centering
\includegraphics[keepaspectratio,width=5in]{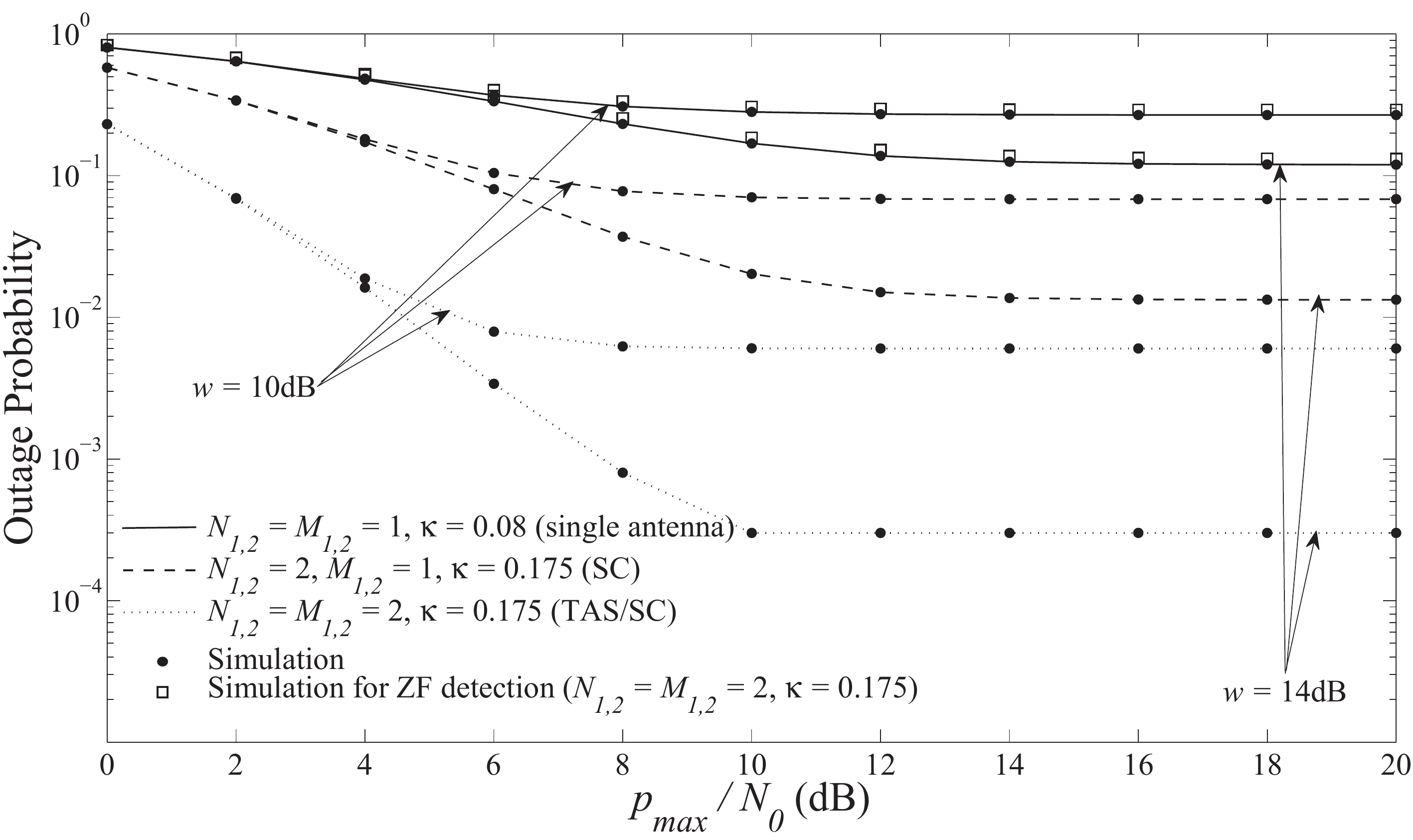}
\caption{Outage probability of the considered cognitive system vs. various signal transmission power regions, when $\mathcal{T}_{\text{P}}=2$, $\gamma_{\text{th}}=3$dB, $\overline{y}=1$dB, and $\overline{x}=4$dB.}
\label{fig2}
\end{figure}

\begin{figure}[!t]
\centering
\includegraphics[keepaspectratio,width=5in]{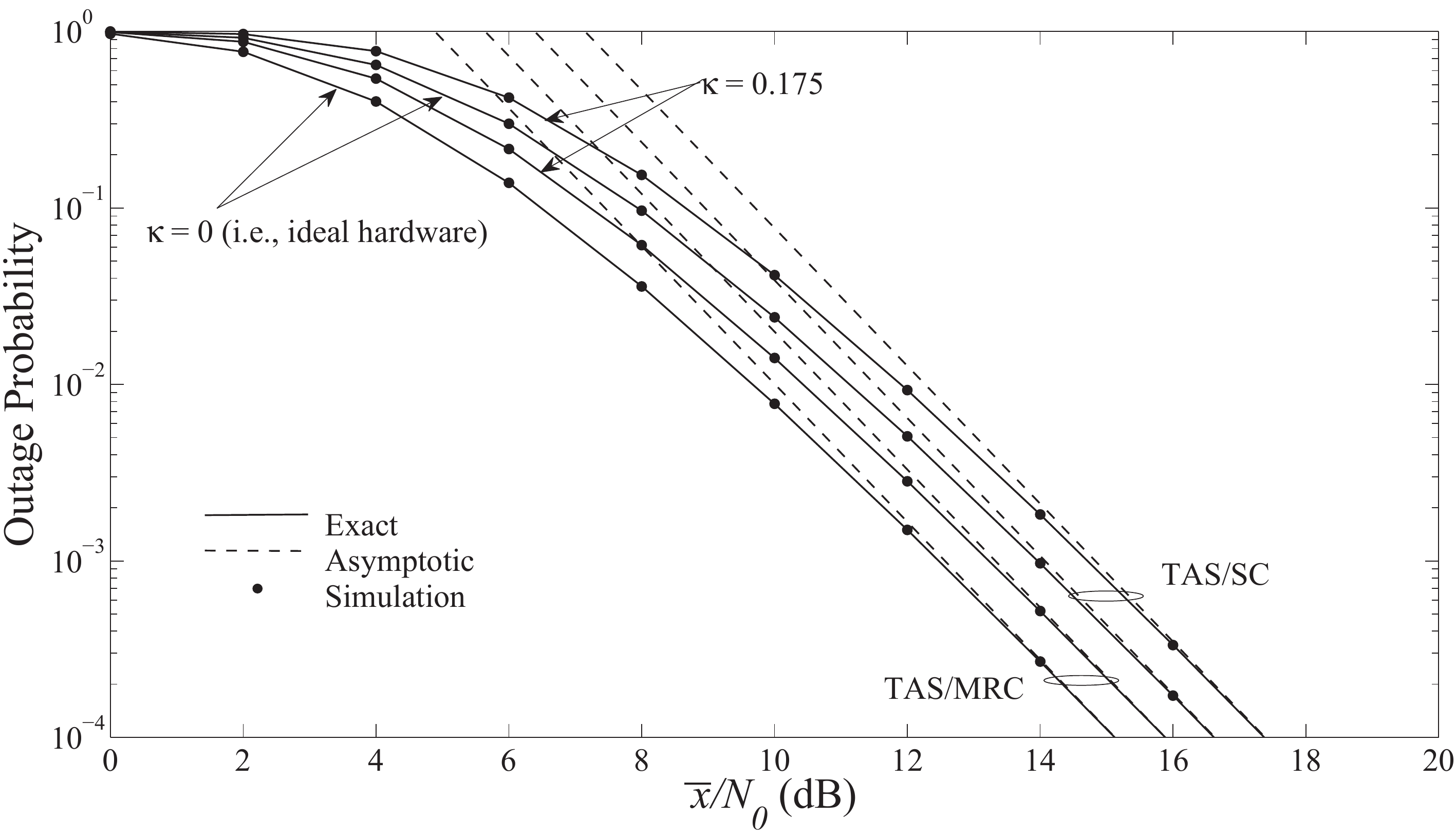}
\caption{Outage probability vs. various channel power fading gain regions, where each secondary node is equipped with 2 antennas ($N_{\left\{1,2\right\}}=M_{\left\{1,2\right\}}=2$). Also, $\mathcal{T}_{\text{P}}=1$, $\gamma_{\text{th}}=w=6$dB, $p_{\max}=10$dB and $\overline{y}=1$dB.}
\label{fig3}
\end{figure}

\begin{figure}[!t]
\centering
\includegraphics[keepaspectratio,width=5in]{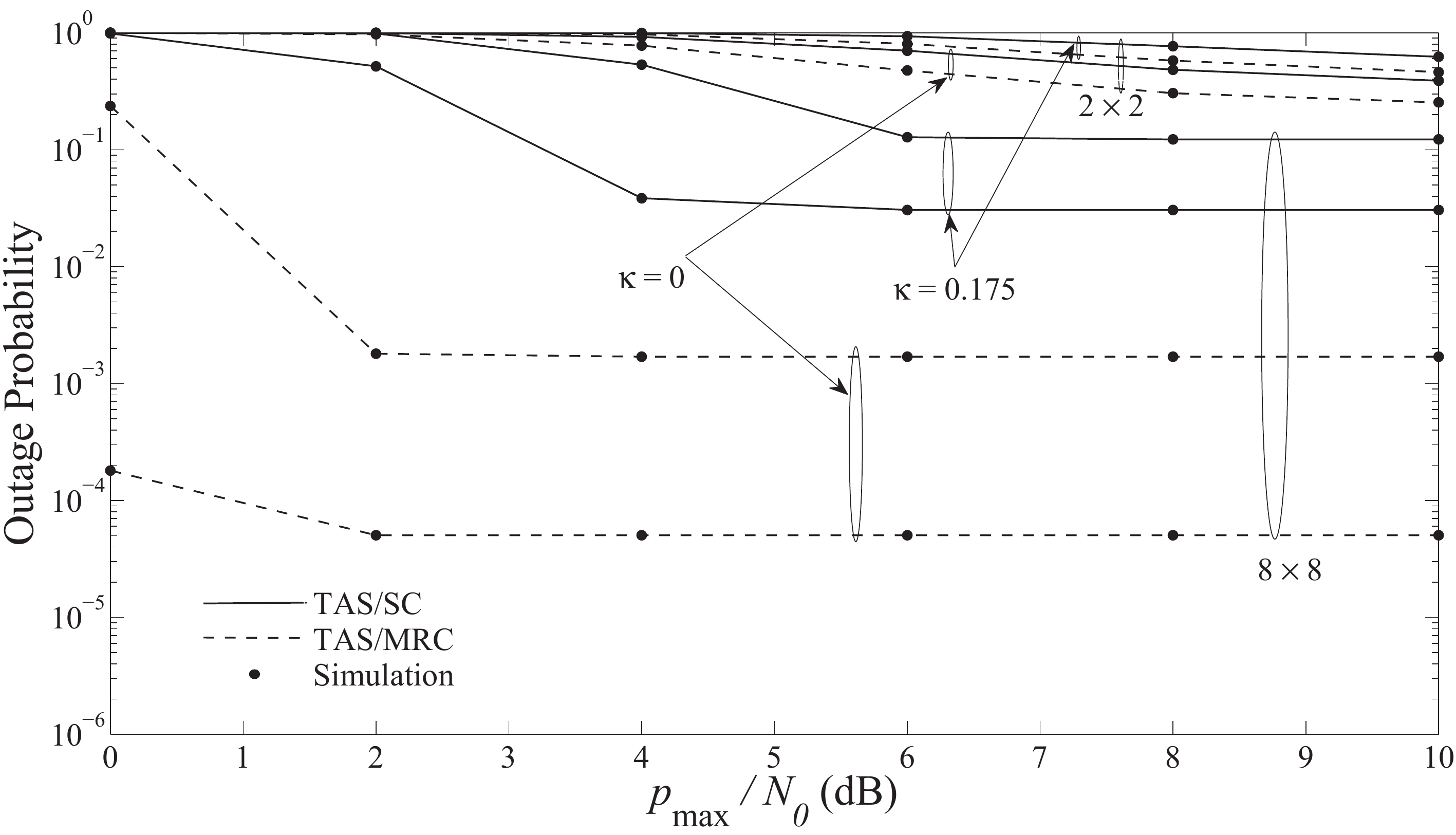}
\caption{Outage probability vs. various signal transmission power regions, when $\mathcal{T}_{\text{P}}=1$, $\gamma_{\text{th}}=w=8$dB and $\overline{y}=\overline{x}=1$dB and when $M_{\left\{1,2\right\}}=N_{\left\{1,2\right\}}=2$ (i.e., $2\times 2$) or $M_{\left\{1,2\right\}}=N_{\left\{1,2\right\}}=8$ (i.e., $8\times 8$).}
\label{fig4}
\end{figure}
In this section, the theoretical results are presented and compared with Monte-Carlo simulations. For ease of tractability and without loss of generality, we assume symmetric levels of impairments at the transceiver, i.e., an equal hardware quality at the transmitter and receiver\footnote{This is the most efficient solution in terms of performance improvement at the transceiver, as indicated in \cite{ref1,ref7}.} of each node, i.e., $\kappa$(=$\kappa_{T}=\kappa_{R}$). Also, assume identical statistics for each hop, e.g., $w,n,\overline{y}$ and $\overline{x}$.

Notably, one may observe from Fig. \ref{fig2} that it is preferable to enable multiple antennas with non-ideal (e.g., low cost) transceivers rather than single-antenna transceivers with ideal hardware. Hence, spatial diversity seems to overcome hardware impairments at the transceiver, even if the rather suboptimal SC (as compared to the performance of TAS/SC or TAS/MRC) scenario is used. Also, the value of interference threshold plays an important role to the outage probability, since it dramatically affects the floor on outage occurrence. The performance of a typical zero forcing (ZF) detection scheme is also presented for cross-comparison reasons. It can be seen that ZF is inferior to TAS/SC and SC schemes due to the reduced spatial diversity gain.

Moreover, the tightness of the asymptotic curves in moderately medium-to-high channel power gain regions is depicted in Fig. \ref{fig3}. It can be readily seen that the diversity orders of the considered schemes remain unaffected from the impact of hardware impairments. The same argument holds for the performance difference between these schemes. Finally, the superiority of TAS/MRC against TAS/SC, in both cases of ideal and non-ideal hardware, is verified in Fig. \ref{fig4}. Obviously, the impact of hardware impairments plays a key role to the overall performance of both schemes, regardless of the number of transmit/receive antennas.

\section{Conclusion}
Current study provides some new results satisfying performance requirements of practical cognitive multiple-antenna DF relaying systems with non-ideal hardware: a) New straightforward and exact closed-form outage performance expressions are derived; b) The selection of equal number of antennas at the transceiver of each link is the most optimal solution regardless of the amount of hardware impairments; and c) TAS/MRC always outperforms TAS/SC, while such a performance difference is irrespective of hardware quality.

\end{document}